\documentstyle[aps,prl,preprint,epsfig]{revtex}
\tighten
\draft
\begin{document}

\renewcommand{\baselinestretch}{1.}
\renewcommand{\theenumi}{(\alph{enumi})}
\title{\boldmath First Measurement of the Branching Fraction of the Decay
$\psi(2S)\rightarrow\tau^{+}\tau^{-}$ }

\author{
J.~Z.~Bai,$^1$   Y.~Ban,$^5$      J.~G.~Bian,$^1$
I.~Blum,$^{12}$ 
G.~P.~Chen,$^1$  H.~F.~Chen,$^{11}$  
J.~Chen,$^3$ 
J.~C.~Chen,$^1$  Y.~Chen,$^1$ Y.~B.~Chen,$^1$
% ...
\\ (BES Collaboration)}

\address{
$^1$Institute of High Energy Physics, Beijing 100039, People's Republic of
 China\\
$^2$California Institute of Technology, Pasadena, California 91125\\
$^3$Colorado State University, Fort Collins, Colorado 80523\\
$^4$Hangzhou University, Hangzhou 310028, People's Republic of China\\
$^5$Peking University, Beijing 100871, People's Republic of China\\
$^6$Shandong University, Jinan 250100, People's Republic of China\\
$^7$Shanghai Jiaotong University, Shanghai 200030, People's Republic of China\\
$^8$Stanford Linear Accelerator Center, Stanford, California 94309\\
$^9$University of Hawaii, Honolulu, Hawaii 96822\\
$^{10}$University of California at Irvine, Irvine, California 92717\\
$^{11}$University of Science and Technology of China, Hefei 230026,
People's Republic of China\\
$^{12}$University of Texas at Dallas, Richardson, Texas 75083-0688}

\date{Received 27 October 2000}
\maketitle

\begin{abstract}
 The branching fraction of the $\psi(2S)$ decay into $\tau^{+} \tau^{-}$ has
been measured for the first time using the BES detector at the Beijing
Electron-Positron Collider. The result is
$B_{\tau\tau}=(2.71\pm 0.43 \pm 0.55)\times 10^{-3}$, where the first error is
statistical and the second is systematic. This value, along with those for
the branching fractions into $e^+e^-$ and $\mu^+\mu^-$ of this 
resonance, satisfy well the relation predicted by the sequential lepton 
hypothesis. Combining all these values with the leptonic width of the resonance, 
the total width of the $\psi(2S)$ is determined to be (252$\pm$37) keV.
\end{abstract}

\par
The $\psi(2S)$ provides a unique opportunity
to compare the three lepton generations by studying the leptonic decays
$\psi(2S)\rightarrow e^+e^-$, $\mu^{+}\mu^{-}$, and $\tau^{+}\tau^{-}$. 
The sequential lepton hypothesis leads to a relationship
between the branching fractions of these decays, $B_{ee}$, $B_{\mu\mu}$, and
$B_{\tau\tau}$ given by
\begin{equation}
\frac{B_{ee}}{v_{e}(\frac{3}{2}-\frac{1}{2}v_{e}^{2})}  =
\frac{B_{\mu\mu}}{v_{\mu} (\frac{3}{2}-\frac{1}{2}v_{\mu}^{2})}=  
\frac{B_{\tau\tau}}{v_{\tau} (\frac{3}{2}-\frac{1}{2}v_{\tau}^{2})}
\label{e1}
\end{equation}
with $v_{l}=(1-\frac{4m_{l}^{2}}{M_{\psi(2S)}^2})^{\frac{1}{2}}$, $l=e$, $\mu, 
\tau$ . Substituting mass values for the leptons and the $\psi(2S)$ gives
\begin{equation}
B_{ee}\simeq B_{\mu\mu}\simeq \frac{B_{\tau\tau}}{0.3885} \equiv B_{ll}
\label{e2}
\end{equation}
Previous experiments have provided measurements of $B_{ee}$ and $B_{\mu\mu}$ 
for the $\psi(2S)$~$\cite{Bee}\cite{Bmm}$
. We present here the first  measurement of
$B_{\tau\tau}$ for the $\psi(2S)$ and compare it to the existing
measurements of $B_{ee}$ and $B_{\mu\mu}$ for this resonance. Combining these
values with previous result for the leptonic width of this 
resonance~$\cite{Lwid1}\cite{Lwid2}$, we determine the total width of the 
$\psi(2S)$.
\par
The data were taken with the Beijing Spectrometer (BES) 
at the Beijing Electron--Positron Collider (BEPC). 
BES, a general--purpose magnetic detector, has been described in detail 
elsewhere~$\cite{BES}$. Briefly, a central drift chamber surrounding the
beam pipe is used for trigger purposes. The main drift-chamber system, 
measures the momentum of charged tracks over $85\%$ of
the $4\pi$ solid angle with a resolution of $\sigma_{p}/p=1.7\%\sqrt{1+p^2}$
($p$ in GeV/$c$). Complementary measurements of specific 
ionization ($d$E/$dx$) and time of flight are used for particle 
identification.   The $d$E/$dx$ resolution for minimum ionizing particles 
is 9\%.  Scintillation counters measure the time-of-flight of charged
particles over $76\%$ of 4$\pi$ with a resolution of 330~ps for Bhabha
events and 450~ps for hadrons. A
cylindrical twelve-radiation-length Pb/gas electromagnetic calorimeter
operating in self-quenching streamer mode and covering $80\%$ of $4\pi$ provides
an energy resolution of $\sigma_{E}/E=22\%/\sqrt{E}$ ($E$ in GeV)
and spatial resolutions of $\sigma_{\phi}=7.9$~mrad, 
and $\sigma_{z}=3.6$~cm. Endcap time-of-flight 
counters and shower counters are not used in this analysis. A conventional
solenoid encloses the calorimeter, providing a 0.4T field. The outermost 
component is a three-layer iron flux return instrumented for 
muon identification which
yields spatial resolutions of $\sigma_{z}=5$~cm and $\sigma_{r\phi}=3$~cm 
over $68\%$ of
$4\pi$ for muons with momenta greater than 550~MeV/$c$.
\par
  This analysis is based on a total integrated luminosity of about 
6.1 pb$^{-1}$ at a center-of-mass energy corresponding to 
the $\psi(2S)$ resonance 
with an uncertainty of 0.29~MeV. The spread in the center-of-mass energy 
of the collider is $\Delta=(1.4\pm 0.1)$~MeV.  The data, a total of 3.96
million $\psi(2S)$ events, were collected in two separate running periods.  
Because of the difference in running conditions of the detector in the
two periods, the two distinct data sets, I and II, are analyzed separately.
\par
 The $\tau^+\tau^-$ events are identified by requiring that 
one $\tau$ decays via $e\nu\overline{\nu}$ and the other via
$\mu\nu\overline{\nu}$. To select
candidate $\tau^+\tau^-$ events, it is first required that exactly
two oppositely charged tracks be
well reconstructed. For each
track, the point of closest approach to the beam line must have 
$|r|<1.5$~cm, and $|z|<15$~cm,
where $z$ is measured along the beam line from
the nominal beam crossing point.
The acolinearity angle, $\theta_{acol}$, defined as the angle between
the outgoing charged tracks, is required to satisfy
$10^{\circ}\leq\theta_{acol}\leq 170^{\circ}$ to reject Bhabhas, muon pairs
and cosmic rays. 
The acoplanarity angle, $\theta_{acop}$, defined as
the angle between the planes defined by the beam direction and the momentum
vector of each charged track, is required to satisfy 
$\theta_{acop}\geq 20^{\circ}$ to suppress radiative Bhabhas and radiative
muon pairs.
Furthermore, each track is required to satisfy $|\cos\theta|\leq 0.65$, 
where $\theta$ is the polar angle, to ensure that it is contained within 
the fiducial region of the barrel electromagnetic calorimeter.
\par
 Next, it is required that the transverse momentum of each charged track be
above the 70~MeV/$c$ minimum needed to traverse the barrel time-of-flight
counter and reach the outer radius of the calorimeter in the 0.4~Tesla
magnetic field. In addition, the momentum must be less than
the maximum kinematically allowed value for a $\tau$ decay at the given 
c.m. energy within a tolerance of 3 standard deviations in momentum resolution.
\par
The search for $\tau^+\tau^-$ production events is restricted to final states
which do not contain
$\pi^{0}$'s or $\gamma$'s. Consequently, there should be no isolated photon
present in the calorimeter, which is defined as an electromagnetic shower
having energy $>60$~MeV and a separation from the nearest charged track of
at least $12^{\circ}$. 
\par
 A particle identification procedure is applied to the selected events. 
 Using the information provided by the main drift chamber
($d$E/$dx$), the scintillation counters (time-of-flight), the electromagnetic
calorimeter (shower energy), we define Xse as the $d$E/$dx$ separation, Tse as the
TOF separation and Sse as the shower energy separation, all assuming 
the electron hypothesis. Here, separation means \{ (measured value -
expected value)/resolution \}.
Then, to identify a track as a electron we required $-4 \leq Tse \leq 0.5, 
-1 \leq Xse \leq 2, -4 \leq Sse \leq 4$ if its momentum is less
than 0.35GeV ; $-4 \leq Tse \leq 1.5, -2 \leq Xse \leq 2, 
-1.5 \leq Sse \leq 4$ if its momentum is between 0.35GeV and 0.7GeV ; or
$-4 \leq Tse \leq 4, -1.5 \leq Xse \leq 2, -2 \leq Sse \leq 4$ if its
momentum is greater than 0.7GeV. A track is assigned as a muon if there are
at least 2 hits in the muon counters.
\par
 The same requirements are applied to 5 million events from a control sample
taken at the $J/\psi$ energy to estimate the expected contributions of 
backgrounds $n_{bg}$ to be subtracted from the selected $e\mu$ events $n_{e\mu}$.
Only one event meets the criteria for the $e\mu$ topology, which corresponds to a 
background of 0.24 events for data set I and 0.49 events for data set II.
A Monte Carlo study on the two--photon process has also been performed;
its contamination is estimated to be negligible.
\par
To obtain the number of resonant $\tau$-pair events, the QED contribution
including the interference effect is subtracted from the total number of 
$\tau^{+}\tau^{-}$ events. $B(\tau\tau)$ is calculated from
\begin{equation}
B(\tau\tau)=\frac{(n_{e\mu}-n_{bg})/B\epsilon_{trig}\epsilon_{d}-\sigma_{Q+I}
{\cal L}} {N_{\psi(2S)}}.
\label{e5} 
\end{equation}
Here 
B is the fraction of $\tau^+\tau^-$ events yielding the $e\mu$ topology,
  which is equal to 0.06194~$\cite{PDG}$; 
$\epsilon_{trig}$ is the trigger efficiency, which for $e\mu$ events within
  fiducial volume is estimated to be approximately $100\%$;
$\epsilon_d$ is the detection efficiency, which is determined by
  using $4 \times 10^{5}$ Monte Carlo-simulated events that are generated by 
  KORALB~$\cite{KORALB}$. The results are $\epsilon_d=14.49\%$ for data set I 
  and $14.39\%$ for data set II (the luminosity-weighted average of 
  $\epsilon_d$ for the whole data is $14.42\%$).
The calculated QED $\tau$-pair cross section
  including interference $\sigma_{Q+I}$ is equal to $2.230nb^{-1}$ at the 
  center-of-mass energy corresponding to the $\psi(2S)$ resonance~$\cite{QED}$ ; 
$N_{\psi(2S)}$ is the number of produced $\psi(2S)$ events; and 
${\cal L}$ is the accumulated luminosity at the resonance. 
\par
\begin{figure}[h]
  \center{\mbox{\psfig{file=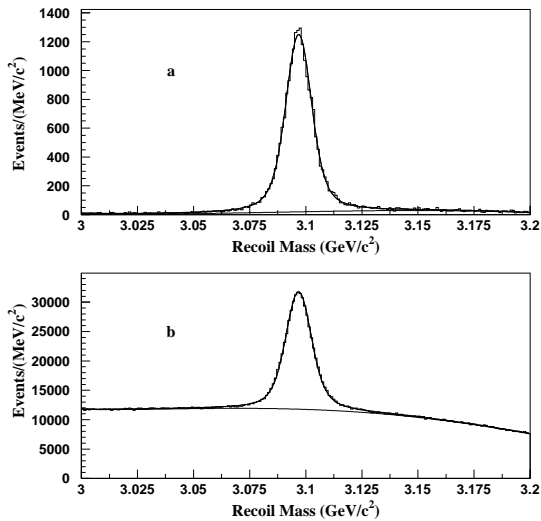,width=7.8cm,height=7.8
   cm}}}
\caption{The mass recoiling against the $\pi^{+}\pi^{-}$ system from
$\psi(2S)\rightarrow \pi^{+}\pi^{-}J/\psi$:  (a) exclusive 
$J/\psi\rightarrow l^+l^-$ events; (b) inclusive $J/\psi$ events.}
\label{fig1}
\end{figure}

The number of produced $\psi(2S)$ events $N_{\psi(2S)}$ is determined from 
a study of inclusive $J/\psi$ produced in $\psi(2S)$ decays in the topology
$\psi(2S) \rightarrow \pi^{+}\pi^{-}J/\psi$~$\cite{Npsip}$.
The number of produced $\pi^{+}\pi^{-}J/\psi$ events,
$N_{\pi\pi J/\psi}$ is
inferred in the recoil from the $\pi^{+}\pi^{-}$ system (Fig. \ref{fig1}).
To estimate $N_{\psi(2S)}$, the PDG value for
$B(\psi(2S) \rightarrow \pi^{+}\pi^{-}J/\psi)=(31.0\pm2.8)\%$~$\cite{PDG}$
is used.
\par
\begin{figure}
  \center{\mbox{\psfig{file=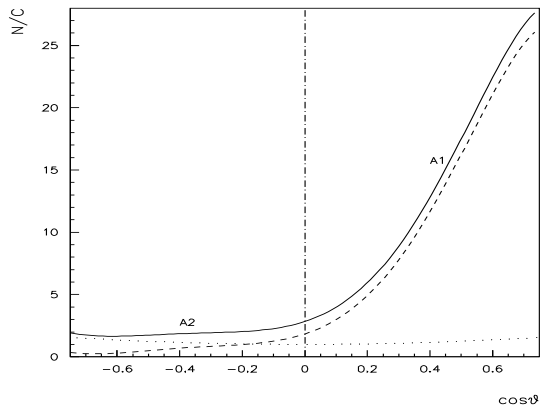,width=7.8cm,height=5.8
   cm}}}
\caption{Schematic of angular distributions of electrons (or
positrons) produced in $e^{+}e^{-} \rightarrow e^{+}e^{-}$ at c.m.
energy=3.686 GeV. The dashed curve represents the Bhabha
scattering (corrected for interference); the
dotted curve represents the resonant $e^{+}e^{-}$ production of the
$\psi(2S)$ and of the $J/\psi$ from $\psi(2S) \rightarrow$ neutral $J/\psi$
decays; the solid curve shows their sum.}
\label{fig2}
\end{figure}

The luminosity ${\cal L}$ is determined by using wide-angle Bhabha events 
at the $\psi(2S)$ in the BES detector and is given by
\begin{equation}
{\cal L}=\frac{N_{QED}}
{\sigma_{QED} \epsilon_{t} \epsilon_{d}},
\label{e6}
\end{equation}
where $N_{QED}$, $\sigma_{QED}$, $\epsilon_{t}$, and
 $\epsilon_{d}$ refer, respectively, to the
observed number of Bhabha events at the $\psi(2S)$, the Bhabha cross section
corrected for interference at the resonance, the
trigger efficiency, and the
detection efficiency for Bhabha events. In order to obtain pure Bhabha
events, the $e^{+}e^{-}$ events from $\psi(2S) \rightarrow e^{+}e^{-}$ as well
as from $\psi(2S) \rightarrow$ neutral $J/\psi$, $J/\psi \rightarrow
e^{+}e^{-}$, should be subtracted from the total number of events. 
These events are symmetric in $\cos\theta$ while the Bhabha events are
asymmetric in $\cos\theta$. 
Using the $\cos\theta$ distribution for $e^{+}e^{-}$ production
relative to $\cos\theta = 0$ as shown in Fig. \ref{fig2}, a relation for the number of
Bhabha events can be obtained
\begin{equation}
N_{QED}=\frac{A_{1}-A_{2}}
{1-2\alpha},
\label{e7}
\end{equation}
where $A_{1}$ and $A_{2}$ are the total number of $e^{+}e^{-}$ events
found from the area under the solid curve on the right side and left side
relative to $\cos\theta = 0$, respectively, and $\alpha$ is the fraction
of Bhabha events on the left side, which is determined by a Monte Carlo
simulation.
\par
The results of this measurement are summarized in Table \ref{Npsip} and 
Table \ref{Ntt}. Combining the results
from the two different running periods, the branching
fraction of the $\psi(2S)$ decaying into $\tau^{+}\tau^{-}$ is calculated to be
\begin{equation}
B_{\tau\tau}=(2.71 \pm 0.43 \pm 0.55) \times 10^{-3},
\label{e8}
\end{equation}
where the first error is statistical and the second is systematic. The
overall relative systematic error of $20.2\%$ includes contributions from the
luminosity ${\cal L}$ $(3.1\%)$;
the number of $\psi(2S)$ events, $N_{\psi(2S)}$ $(9.1\%)$;
the selection criteria for $e\mu$ topology $(11.3\%)$;
and the calculated values of $\sigma_{\rm QED}$ 
due to uncertainties in the c.m. energy scale and the spread
in c.m. energy $(10.8\%)$. 

In Table \ref{lepton}, we summarize the existing measurements
of the leptonic decays of the $\psi(2S)$. Our value of $B_{\tau\tau}$,
corrected by a factor of 0.3885, as indicated in Eq.(\ref{e2}), agrees with 
the values of $B_{ee}$ and $B_{\mu\mu}\cite{PDG}$. Assuming 
lepton universality, the average value $B_{ll}$ is
determined to be $(8.4\pm 1.0)\times 10^{-3}$. The leptonic width
$(\Gamma_{ee})$ of the $\psi(2S)$ has been determined to be 
$(2.12\pm 0.18)$~keV $\cite{PDG}$. 
From the relationship $\Gamma_{tot}=\Gamma_{ee}/B_{ll}$ we find
$\Gamma_{tot}=(252\pm 37)$~keV, which is consistent with the direct 
measurement value ($306\pm 39)$~keV by E760~$\cite{E760}$ within about 
one standard deviation.
\par
 In conclusion, we have measured $B_{\tau\tau}$ for the $\psi(2S)$. This
result, along with the previous data of $B_{ee}$ and $B_{\mu\mu}$, 
satisfy well the relation predicted by the sequential lepton hypothesis. 
Combining these values we have calculated the total width for this resonance. 
\par
 We gratefully acknowledge the efforts of the staffs of the BEPC accelerator
and the computing center at  the Institute of High Energy Physics, Beijing. 
We thank 
B.~N.~Jin for his contribution to the calculation of QED cross sections.
This work is supported in part by the National Natural
 Science Foundation of China under Contract No. 19290400 and the Chinese
 Academy of Sciences under contract No. H-10 and E-01 (IHEP),
 and by the Department of Energy under Contract Nos.
 DE-FG03-92ER40701 (Caltech), DE-FG03-93ER40788 (Colorado State University),
 DE-AC03-76SF00515 (SLAC), DE-FG03-91ER40679 (UC
 Irvine), DE-FG03-94ER40833 (U Hawaii), DE-FG03-95ER40925 (UT Dallas).


TABLES\\

%

\begin{table}
\caption{Numbers used to calculate $B_{\tau\tau}$. 
The first error is statistical and the second is systematic.}
\begin{tabular}{ccccccc}
Data set & $n_{e\mu}$ & $n_{bg}$ & $\epsilon_{d}$ & $\cal{L}$(pb$^{-1}$) 
& $N_{\pi\pi J/\psi}(10^6)$ & $N_{\psi(2S)}(10^6)$  \\ \hline
I & 77 & 0.27 & 0.1449 & $2.123\pm0.015\pm0.051$ 
& $0.4293\pm0.0017\pm0.0076$ & $1.385\pm0.005\pm0.127$ \\
II & 140 &0.49 & 0.1439 & $3.929\pm0.019\pm0.098$ 
& $0.7980\pm0.0023\pm0.0092$ & $2.574\pm0.007\pm0.234$ \\
Total & 217 & 0.76 &0.1442 & $6.052\pm0.024\pm0.149$ 
& $1.227\pm0.003\pm0.017$ & $3.959\pm0.009\pm0.362$ \\
\end{tabular}
\label{Npsip}  
\end{table}

\vspace{15mm}

\begin{table}
\caption{Branching fraction $B_{\tau\tau}/B_{\pi^{+}\pi^{-}J/\psi}$ and
final branching ratio $B_{\tau\tau}$. 
}
\begin{tabular}{ccc}
Data set &  $B_{\tau\tau}/B_{\pi^{+}\pi^{-}J/\psi} (10^{-3})$
&$B_{\tau\tau} (10^{-3})$ \\ \hline
I & $8.89\pm2.35\pm1.61$ & $2.76\pm0.73\pm0.56$ \\
II & $8.63\pm1.72\pm1.63$ & $2.68\pm0.53\pm0.56$ \\
Total & $8.73\pm1.39\pm1.57$ & $2.71\pm0.43\pm0.55$ \\
\end{tabular}
\label{Ntt}  
\end{table}

\vspace{15mm}
\begin{table}
\caption{Leptonic branching fractions of the $\psi(2S)$ in $10^{-3}$.}
\begin{tabular}{ccc}
$B_{ee}$ &$B_{\mu\mu}$ & $B_{\tau\tau}/0.3885$ \\ \hline
$8.8\pm1.3$ &$10.3\pm3.5$ &$7.0\pm1.1\pm1.4$ \\
\end{tabular}  
\label{lepton}
\end{table}

\end{document}